\documentclass[10pt,conference]{IEEEtran}
\usepackage[noadjust]{cite}
\usepackage{amsmath,amssymb,amsfonts}
\usepackage{graphicx}
\usepackage{textcomp}
\usepackage{xcolor}
\usepackage[linesnumbered,lined,commentsnumbered]{algorithm2e}
\usepackage{adjustbox}

\makeatletter
\renewcommand{\@algocf@capt@plain}{above}
\makeatother

\SetAlFnt{\small}

\usepackage{flushend}
\usepackage{balance}

\ifCLASSOPTIONcompsoc
\usepackage[caption=false,font=normalsize,labelfont=sf,textfont=sf]{subfig}
\else
\usepackage[caption=false,font=footnotesize]{subfig}
\fi

\usepackage{pgfplots}
\pgfplotsset{compat=1.8}
\usepgfplotslibrary{statistics}

\usepackage{multirow}

\def\BibTeX{{\rm B\kern-.05em{\sc i\kern-.025em b}\kern-.08em
    T\kern-.1667em\lower.7ex\hbox{E}\kern-.125emX}}

\usepackage{algpseudocode}  
\begin{document}

\title{Recommending Comprehensive Solutions for Programming Tasks by Mining Crowd Knowledge}

\author{
\IEEEauthorblockN{Rodrigo F. G. Silva\IEEEauthorrefmark{1},
Chanchal K. Roy\IEEEauthorrefmark{2},
Mohammad Masudur Rahman\IEEEauthorrefmark{2},\\
Kevin A. Schneider\IEEEauthorrefmark{2},
Klerisson Paixao\IEEEauthorrefmark{1},
Marcelo de Almeida Maia\IEEEauthorrefmark{1}}

\IEEEauthorblockA{\IEEEauthorrefmark{1}Federal University of Uberl{\^a}ndia, Uberl{\^a}ndia (MG), Brazil\\ 
\{rodrigofernandes, klerisson, marcelo.maia\}@ufu.br}

\IEEEauthorblockA{\IEEEauthorrefmark{2}University of Saskatchewan, Saskatoon, Canada \\ \{chanchal.roy, masud.rahman, kevin.schneider\}@usask.ca  \vspace{-3em}}

}

\maketitle

\begin{abstract}
Developers often search for relevant code examples on the web for their programming tasks. Unfortunately, they face two major problems. First, the search is impaired due to a lexical gap between their query (task description) and the information associated with the solution. Second, the retrieved solution may not be comprehensive, i.e., the code segment might miss a succinct explanation. These problems make the developers browse dozens of documents in order to synthesize an appropriate solution. To address these two problems, we propose CROKAGE (Crowd Knowledge Answer Generator), a tool that takes the description of a programming task (the query) and provides a comprehensive solution for the task. Our solutions contain not only relevant code examples but also their succinct explanations. Our proposed approach expands the task description with relevant API classes from Stack Overflow Q\&A threads and then mitigates the lexical gap problems. Furthermore, we perform natural language processing on the top quality answers and then return such programming solutions containing code examples and code explanations unlike earlier studies. We evaluate our approach using 97 programming queries, of which 50\% was used for training and 50\% was used for testing, and show that it outperforms six baselines including the state-of-art by a statistically significant margin. Furthermore, our evaluation with 29 developers using 24 tasks (queries) confirms the superiority of CROKAGE over the state-of-art tool in terms of relevance of the suggested code examples, benefit of the code explanations and the overall solution quality (code + explanation).

\end{abstract}

\begin{IEEEkeywords}
Mining Crowd Knowledge, Stack Overflow, Word Embedding
\end{IEEEkeywords}

\vspace{-1.5em}
\section{Introduction} \label{Introduction}

Software developers often search for relevant code examples on the web to implement their programming tasks. Although there exist several Internet-scale code search engines (e.g., Koders, Krugle, GitHub), finding code examples on the web is still a major challenge. Developers generally choose a few important keywords to describe their programming task, and then submit their query to a code search engine (e.g., Koders). Unfortunately, they face two major problems. First, the search is impaired due to a lexical gap between the task description (the query) and the information pertinent to the solution. Their query often does not contain the API references required for the task. Second, the retrieved solution might always not be comprehensive. The retrieved code segment might miss a succinct explanation~\cite{rahman2018effective} or the textual solution might miss a required code segment~\cite{xu2017answerbot}. These problems make the developers browse dozens of search results in order to synthesize an appropriate solution for their task. 

Traditional Information Retrieval (IR)-based code search engines generally do not work well with natural language queries due to a lexical mismatch between the keywords of a query and the available code examples on the web.  Mikolov~et~al.~\cite{mikolov2013distributed} recently employ word embedding technology that captures words' semantics, represents each word using a high-dimensional vector and then estimates the semantic similarity between any two documents despite their lexical dissimilarity. This technique was later used by two recent studies: AnswerBot~\cite{xu2017answerbot} and BIKER~\cite{huang2018api}. They also attempt to address the two aforementioned problems faced by developers. However, these studies are limited in several aspects. AnswerBot's answers do not contain any source code examples. Thus, they are not sufficient enough for implementing a \textit{how-to} programming task. On the other hand, BIKER is able to provide answers containing both explanations and source code. However, BIKER is only able to provide the explanations from official Java SE API documentations. Thus, their explanations are restricted to a limited set of APIs only. 

\begin{figure*}
\centerline{\includegraphics[width=1.0\textwidth]{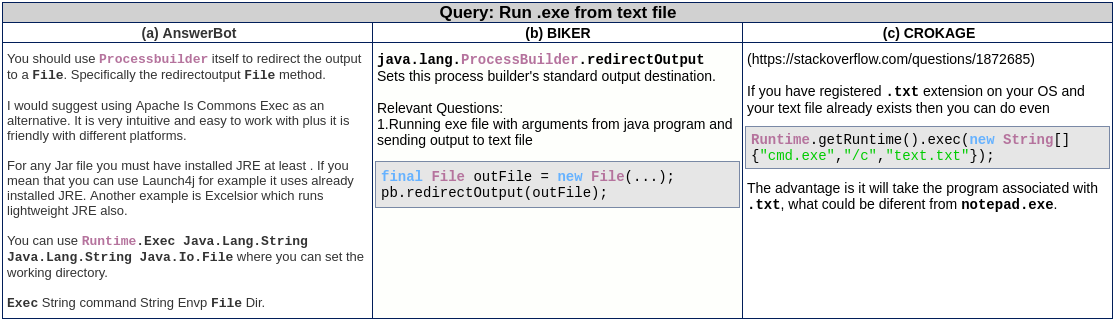}}
\vspace{-1em}
\caption{Programming Solutions from (a) AnswerBot, (b) BIKER, and (c) CROKAGE}
\label{fig:preamblesoanswer1}
\vspace{-1.5em}
\end{figure*}

In this paper, we propose an approach namely CROKAGE (\textbf{Cro}wd \textbf{K}nowledge \textbf{A}nswer \textbf{Ge}nerator) that takes a task description in natural language (the query) and then returns relevant, comprehensive programming solutions containing both code examples and succinct explanations. In particular, we address the limitations of the two earlier approaches \cite{xu2017answerbot,huang2018api}. First, unlike AnswerBot~\cite{xu2017answerbot} (i.e., provides only answer summary texts), we deliver both relevant code segments and their corresponding explanations. Second, we provide succinct code explanations written by human developers whereas BIKER~\cite{huang2018api} returns only generic explanations extracted from official API documentations. CROKAGE first employs word embeddings~\cite{mikolov2013distributed} to overcome the lexical gap between the query and the code example and then expands the task description (the query) with relevant API classes from Stack Overflow. Then it constructs a multi-factor retrieval mechanism to fetch from Stack Overflow the most relevant code examples to the target programming task. Furthermore, CROKAGE uses natural language processing to compose a succinct explanation for each of the suggested code examples.

We evaluate our approach in two different ways. First, we compare our performance in code example suggestion with six baselines including BIKER~\cite{huang2018api}, TF-IDF~\cite{salton1988term} and BM25~\cite{robertson1994some}. For this, we use four different metrics and show that our approach outperforms all baselines by a statistically significant margin. We construct our ground truth by manually analyzing 6,224 answers from Stack Overflow against 100 questions (or queries) collected from three popular tutorial sites (KodeJava, JavaDB, Java2s). We select 97 questions containing API classes in their answers, of which 50\% was used for training and 50\% was used for testing. 
Our experiments show that CROKAGE provides relevant programming solutions (code segments + explanations) with 79\% Top-10 Accuracy, 40\% precision, 19\% recall, and a reciprocal rank of 0.46 which are 64\%, 30\%, 18\%, and 36\% higher respectively than those of the state-of-art, BIKER~\cite{huang2018api}. Second, we conduct a user study with 29 developers using 24 programming tasks. Our findings suggest that solutions from CROKAGE are more effective than the ones of BIKER~\cite{huang2018api} in terms of relevance of the suggested code examples, benefit of the code explanations and the overall solution quality (code + explanation).

Thus, the main contributions of this paper are as follows:

\begin{itemize}

\item A novel approach that suggests programming solutions containing both code and explanations against tasks written in natural language texts by harnessing the crowd knowledge stored in Stack Overflow.

\item We perform an empirical evaluation on the suggestion of relevant code examples using 97 programming tasks containing API classes in their answers and a comparison with the state-of-the-art study~\cite{huang2018api}. Our approach achieves significant improvements over six baselines and outperforms the state-of-art in retrieving relevant and comprehensive programming solutions. 

\item A ground truth and benchmark dataset of 6,224 answers against 100 Java tutorial questions constructed by two professional developers after spending 87 man hours.

\item A replication package\footnote{https://github.com/muldon/CROKAGE-replication-package} containing CROKAGE's prototype, detailed results of our user study and our used dataset for replication or third party reuse.

\end{itemize}

The rest of this paper is structured as follows.  Section~\ref{sec:motivation} shows the motivation of our work and compare our approach with two state-of-art tools. Section~\ref{TheMethodology} describes the technical details of CROKAGE. Section~\ref{experiments} reports the experimental methods and the obtained results. Section~\ref{sec:threats} discusses the threats to validity of our experiments. Section~\ref{relatedWord} reviews the related works. Finally, Section~\ref{conclusion} concludes the paper.

\section{Motivating Examples}\label{sec:motivation}

Let us consider a use-case scenario where a developer is looking for a solution to the query: \textit{``run .exe from text file"}. Figure~\ref{fig:preamblesoanswer1} presents three solutions from three different approaches: AnswerBot~\cite{xu2017answerbot}, BIKER~\cite{huang2018api} and CROKAGE respectively. The solution proposed by AnswerBot (Fig.~\ref{fig:preamblesoanswer1}-(a)) contains sentences describing the use of several API classes as well as opinions from Stack Overflow users. Despite describing solutions using relevant APIs (e.g., \texttt{Processbuilder, Runtime}), no actual code is provided. That is, the solution is half-baked and thus might not help the developer properly.

On the contrary, BIKER~\cite{huang2018api} provides both code example and corresponding explanation (i.e., Fig.~\ref{fig:preamblesoanswer1}-(b)). Unfortunately we notice two major problems. First, the suggested code does not completely match with the intent of the query. Second, the explanation is limited to only official Java API documentation and thus might fail to explain the functionalities of other external API classes or methods. 

Finally, our approach CROKAGE provides a solution containing (1) a code segment using \texttt{Runtime} API and (2) an associated prose explaining the code (i.e., Fig.~\ref{fig:preamblesoanswer1}-(c)). 
Unlike AnswerBot~\cite{xu2017answerbot}, CROKAGE delivers a solution containing relevant code segment. Unlike BIKER~\cite{huang2018api} (i.e., generates explanation from official API documentation only) CROKAGE delivers a solution containing code segment which is carefully explained and curated by Stack Overflow users. It also should be noted that unlike BIKER, our explanations are much more generic, informative and not restricted to standard Java APIs. Thus, our solution has a much more potential than the existing alternatives (i.e., AnswerBot~\cite{xu2017answerbot} and BIKER~\cite{huang2018api}) as also confirmed by the user study (Section~\ref{userStudy}).

\section{Proposed Approach} \label{TheMethodology}

\begin{figure*}
\centerline{\includegraphics[width=1.0\textwidth]{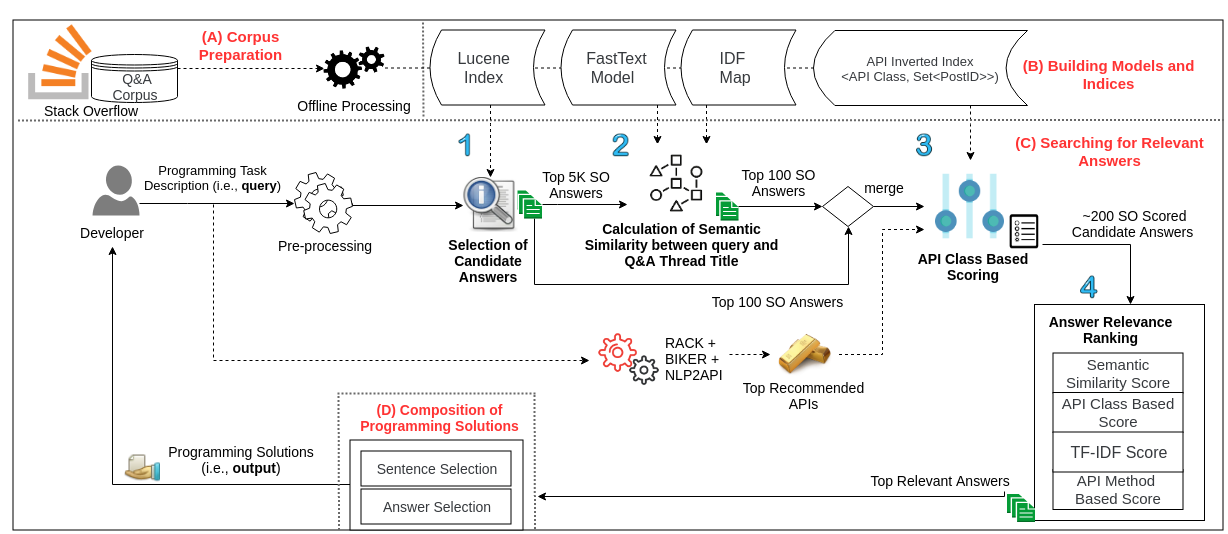}}
\vspace{-1em}
\caption{Schematic diagram of CROKAGE - (A) Corpus Preparation, (B) Building Models and Indices, (C) Searching for Relevant Answers, and (D) Composition of Programming Solutions}
\label{fig:approach}
\vspace{-1.5em}
\end{figure*}

Figure~\ref{fig:approach} shows the schematic diagram of our proposed approach CROKAGE. It has four different stages. We first prepare the corpus using Q\&A threads from Stack Overflow (Section \ref{corpus}), and then construct several models (e.g., \textit{FastText} model) and indices (Section \ref{models}). Then these models and indices are employed to retrieve the relevant answers from the corpus against a programming task description (Section \ref{searchingforrelatedcontentinthecrowd}). Finally, the top quality answers are used to compose and suggest the programming solutions for the task (Section \ref{answerComposition}). We discuss each of these stages as follows:

\subsection{Corpus Preparation} \label{corpus}

In order to deliver appropriate solutions from Stack Overflow against a given task description (i.e., query), we need to construct domain specific knowledge base. We collect a total of 3,889,303 questions and answers related to ``Java" from Stack Overflow Q\&A site\footnote{https://archive.org/details/stackexchange - dump published in June 2018}. We use \textit{Jsoup}~\cite{Jsoup} to parse these questions and answers and then separate texts and code using \texttt{<code>} and \texttt{<pre>} tags. We remove all punctuation symbols, stop words\footnote{https://bit.ly/1Nt4eMh}, small words (i.e., size lower than two) and numbers from them. We save these processed versions of Q\&A threads alongside the original versions which are later used to compose programming solutions (Section \ref{answerComposition}).  

\subsection{Building Models and Indices} \label{models}

\textbf{Construction of Lucene Index}: We first construct a Lucene index by considering all Java related questions and answers from Stack Overflow Fig.~\ref{fig:approach}-(B)). We make sure that each answer from each Q\&A thread contains one or more code segments. Similar to Delfim~et~al.~\cite{delfim2016redocumenting}, we capture the pre-processed version of each question-answer pair as a document within a large document corpus. We then build an index with Lucene~\cite{Lucene} using all these documents. This index is later used to retrieve the most relevant answers against each query (i.e., task description).

\textbf{Construction of FastText Model}: 
Task description (the query) from a developer might always not contain the required API references. Hence, the Lucene-based approach above might fail to retrieve the relevant answers due to lexical gap issue. In order to overcome the lexical gap issue between the query and answers from Stack Overflow, we employ a word embedding model namely \textit{FastText}~\cite{bojanowski2016enriching}. In this model, each word is represented as a high dimensional vector and similar words have similar vector representations. We learn the vector representation of each word of our vocabulary using \textit{FastText}. We first combine the pre-processed contents of \textit{title, body texts} and \textit{code segments} from each of the Q\&A threads into a file and then employ \textit{FastText} to learn the vector representation of each word. We customize these parameters: vector size=100, epoch = 10, minimum size = 2 and maximal size = 5 whereas the other parameters remain default. Please note that we do not perform stemming on the texts since \textit{FastText} is able to look for subwords. Besides, the impact of stemming over source code is found to be controversial~\cite{hill2012use}. Once the model is built (i.e., Fig.~\ref{fig:approach}-(B)), we use it as a dictionary for mapping between the words of our vocabulary and their respective vector representations.

\textbf{Construction of IDF Map}: Inverse Document Frequency (IDF) has been often used for determining the importance of a term within a corpus~\cite{huang2018api,ye2016word}. IDF represents the inverse of the number of documents containing the word. That is, more frequent words across the corpus carry less important information than infrequent words. Our vocabulary contains a total of 1,118,667 distinct words. We calculate the IDF of each word and build an IDF Map (i.e., Fig.~\ref{fig:approach}-(B)) that points each word to its corresponding IDF value. The IDF value of each word is later used as a weight during the calculation of embedding similarity (a.k.a., semantic similarity). 

\textbf{Construction of API Inverted Index}: 
We also build an API inverted index that maps API classes to their corresponding answers from Stack Overflow (i.e., Fig.~\ref{fig:approach}-(B)). To build this index, we first select all the answers containing code (i.e., containing tags \texttt{<pre><code>}), and extract code elements from them using  \textit{Jsoup}~\cite{Jsoup}. Then we identify the API classes from them using appropriate regular expressions. We then build an inverted index where each API class is associated with the IDs of all answers containing that class. We notice that many classes have a low frequency (i.e., lower than five), which originally come from dummy examples submitted by users to explain a very specific scenario.  (e.g., \textit{``You can use @Qualifier on the \texttt{OptionalBean} member"}\footnote{https://stackoverflow.com/questions/9416541} where \texttt{OptionalBean} is a class created by the user to illustrate a scenario). Thus we discard the API classes with low frequencies from our API inverted index. We believe that such API classes could be less appropriate for our problem contexts.

\subsection{Searching for Relevant Answers} \label{searchingforrelatedcontentinthecrowd}

Once the models and indices are built, we use them in searching for relevant answers for a given query. 
Our search component works in four stages as shown in  Figure~\ref{fig:approach}. In order to search for relevant answers, CROKAGE first loads the models and indices (Section \ref{models}) and then pre-processes the task description (i.e., query) with standard natural language pre-processing. Then CROKAGE navigates through the four stages as follows. 

\subsubsection{Selection of Candidate Answers}\label{bm25filter}

Given a pre-processed query and our Lucene index constructed above, CROKAGE uses BM25~\cite{robertson1994some} function to determine the lexical relevance of each answer from the pre-loaded index as follows:

\vspace{-0.9em}
{\small
\begin{equation}
\begin{aligned}
lexicalSim(A,Q)={} & \sum_{i=1}^{n} idf(q_i) \\ \ast
                   & \frac{f(q_i,A)  * (k+1)}{f(q_i,A)+k*(1-b+ b*\frac{|A|}{avgdl})}
\end{aligned}
\end{equation}
}
\vspace{-0.5em}

 \noindent where $|A|$ is the length of the answer $A$ in words, $f(q_i,A)$ is keyword $q_i$'s term frequency in answer $A$ and $avgdl$ is the average document length in the index. $k$ and $b$ are two parameters where $k$ controls non-linear term frequency normalization (saturation), and $b$ controls to what degree document length normalizes term frequency values. CROKAGE uses the same values as used by previous works in Software Engineering \cite{Ahasanuzzaman:2016,silva2018duplicate} with best performance. $idf(q_i)$ is the inverse document frequency of keyword $q_i$ and computed as follows:

\begin{equation}
  idf(q_i)=log\frac{N-n(q_i)+0.5}{n(q_i)+0.5}
\label{eq:bm25idf}    
\end{equation}

 \noindent where $n(q_i)$ is the number of documents containing keyword $q_i$ and $N$ is the total number of documents in the index (or corpus). CROKAGE retrieves the top answers sorted by their relevance to the query and save them into two sets: \textit{smallSet} and \textit{bigSet}. We add the top 100 answers into the \textit{smallSet} and the top 5K answers into the \textit{bigSet}. Both sets are used in the next stages.

\subsubsection{Calculation of Semantic Similarity between Query and Q\&A Thread Title}\label{semanticSimTitles}

This stage takes as input a pre-processed query, the \textit{bigSet} from the previous stage, the \textit{FastText} model and the IDF Map. CROKAGE first transforms the pre-processed information from each Q\&A thread (\textit{body texts + title}) and the query into two bag of words namely A and Q respectively. Then CROKAGE computes the asymmetric relevance between A and Q as follows:

\begin{equation}
    asym(A \rightarrow Q) = \frac{  \sum_{ w \in A }{sim(w,Q)} * idf(w) }{\sum_{ w \in A }{idf(w)} }
\label{eq:asymmetricsim}    
\end{equation}

\noindent where $idf(w)$ is the correspondent IDF value of the word $w$, $sim(w,Q)$ is the maximum value of $sim(w,w_Q)$ for every word $w_Q \in Q$, and $sim(w,w_Q)$ is the cosine similarity between $w$ and $w_Q$ embedding vectors. The other asymmetric relevance namely $asym(Q \rightarrow A$) can be calculated by swapping $A$ and $Q$ in Equation~\ref{eq:asymmetricsim}. Thus, the final similarity between the query $Q$ and the answer $A$ is the harmonic mean of the two asymmetric relevance scores as follows: 

\vspace{-0.2em}
{\small
\begin{equation}
    semScore(A,Q) = \frac{2*asym(A\rightarrow Q)*asym(Q\rightarrow A))}{asym(A\rightarrow Q)+asym(Q\rightarrow A)}
\label{eq:asymmetricsimfinal}    
\end{equation}
}
\vspace{-0.2em}

 \noindent CROKAGE computes semantic relevance between the query and the top 5K answers from the \textit{bigSet} and then selects the top 100 relevant answers which are used for the later stages. 

\subsubsection{API Class Based Scoring}\label{apiScoreSec}
Although the answers from the two previous stages are lexically and semantically relevant to a given query, they might still contain noise. Hence, relevant answers need to be promoted over the noise. One possible way is rewarding the answers based on to their APIs. First, we employ three state-of-art API recommendation systems -- BIKER~\cite{huang2018api}, NLP2API~\cite{rahman2018effective} and RACK~\cite{rahman2016rack} and then collect the most relevant API classes for a given query (i.e., task description). Our findings suggest that the combination of three tools provides the best results which justifies our choice for combining their API suggestions (Section \ref{rankingMechanisms}). We then combine the top 100 answers from the \textit{smallSet} and another top 100 answers from the above stage removing duplicate answers. From each answer, we extract the API classes using appropriate regular expressions and store them in a set called \textit{allApis}, after removing the duplicate classes. Next, for each recommended API class ($c \in C$) from the ranked list (RACK+NLP2API+BIKER), we calculate the API class based score for each answer as follows:

\vspace{-1em}
\begin{equation}
   apiScore(A) = \sum_{c \in C} \frac{1}{pos(c) + n}
\label{eq:apiscore}    
\end{equation}
\vspace{-.5em}

\noindent where $n$ is a smoothing factor and $pos$ is the position (starting with zero) of the class \textit{c} within recommended API ranked list that is also found in \textit{allApis}. After careful investigation, we found the best value of $n$ as two. Our goal is to reward answers with more relevant classes and penalize the answers from Stack Overflow with irrelevant classes. 

\subsubsection{Answer Relevance Ranking}\label{relevanceMechanism}

The previous stages deliver a set of around 200 answers from Stack Overflow along with their lexical, semantic and API relevance with a given query (i.e., task description). However, we consider two additional relevance factors -- TF-IDF Score and API Method Based Score  (Fig.~\ref{fig:approach}-(C-4)) as follows:

\textbf{TF-IDF Score:} Although we use BM25, a lexical similarity method for selecting the candidate answers, we employ another lexical similarity method namely TF-IDF (Fig.~\ref{fig:approach}-(C-4)) to determine their relevance against a given query. TF-IDF stands for term frequency (TF) times inverse document frequency (IDF). It can be calculated for each word of a document (query or answer) as follows:

\vspace{-1em}
\begin{equation}
TF-IDF(W) = TF(W) * log_{10}(\frac{N}{df_{w}})
\label{eq:idf}
\end{equation}
\vspace{-1em}

\noindent We determine lexical similarity between the document representing the query and the document representing each answer using their cosine similarity as follows:

\vspace{-1em}
\begin{equation}
\begin{aligned}
tfidfScore(A,Q) {}= & \frac{d_Q \cdot d_A}{|d_Q| \cdot |d_A|} \\
= & \frac{\sum_{1}^{N} tfidf_{ti,dq} \cdot tfidf_{ti,da}}{\sqrt{\sum_{1}^{N} tfidf_{ti,dq}^{2}} \cdot \sqrt{tfidf_{ti,da}^{2}}}
\label{eq:idfscore}
\end{aligned}
\end{equation}
\vspace{-1em}

\noindent where $d_Q$ refers to the query, $d_A$ represents the answer and $tfidf_{ti,dk}$ is the term weight for each word of the document (query or answer).

\textbf{API Method Based Score:} Huang~et~al.~\cite{huang2018api} suggest that if an API method occurs across multiple candidate answers for a given query, it is more likely relevant to the query. CROKAGE rewards the answers containing the most relevant API methods among the candidate answers. For this, CROKAGE selects the methods used in each answer using appropriate regular expressions and identify the most frequent API method. Then CROKAGE assigns each of the answers an API method based score (Fig.~\ref{fig:approach}-(C-4)) as follows:

\vspace{-1em}
\begin{equation}
methodScore(A) = \frac{log_{2}(freq_m)}{10}
\label{eq:methodScore}
\end{equation}
\vspace{-1em}

 \noindent where $freq_m$ is the top method frequency. If the answer does not contain the top method, the score is set to zero.

After calculating the four scores --- \emph{semScore, apiScore, tfidfScore} and \emph{methodScore}, we normalize them and 
combine them in a final score (\emph{factorsScore}) representing the relevance of each answer A to the query Q:

\vspace{-1em}
{\small
\begin{equation}
\begin{aligned}
factorsScore(Q,A)   
 & = semScore \cdot semWeight \\
 & + apiScore \cdot apiWeight  \\
 & + tfidfScore \cdot tfidfWeight  \\
 & + methodScore \cdot methodWeight  
\end{aligned}
\label{composer}
\end{equation}
} 
\vspace{-.5em}

\noindent where \emph{semWeight}, \emph{apiWeight}, \emph{tfidfWeight}, and \emph{methodWeight} are relative weights for each factor. We conduct controlled iterative experiments and employ a set of weights that return the best Top-K Accuracy and MRR (Section \ref{rankingMechanisms}). Once the final score is calculated, we collect the Top-K Stack Overflow answers for the solution composition.

\subsection{Composition of Programming Solutions} \label{answerComposition}

After collecting the most relevant answers for a given query (i.e., task description), CROKAGE uses them to compose appropriate solutions to the desired task, discarding answers without important sentences. CROKAGE adopts two patterns, previously used by Wong~et~al.~\cite{wong2013autocomment}, to identify important sentences based on their POS structure: 

\begin{equation}
\begin{aligned}
    {}& VP << (NP < /NN.?/) < /VB.?/ \\
    & NP ! < PRP [<< VP | \$ VP]
\end{aligned}
\end{equation}

\noindent These patterns ensure that each sentence has a verb which is associated with a subject or an object. The first pattern guarantees that a verb phrase is followed by a noun phrase while the second pattern guarantees that a noun phrase is followed by a verb phrase. They also ensure that a verb phrase is not a personal pronoun. CROKAGE filters important sentences using pseudocode as shown in Algorithm~\ref{algo:pseudocodee}.

\SetKwProg{Fn}{}{}{end}\SetKwFunction{FRecurs}{filterSentences}%
\newcommand{\forcond}{$i=0$ \KwTo $n$}
\begin{algorithm}
\SetAlgoLined
\SetAlgoLongEnd
  \label{algo:pseudocodee}
  \caption{Pseudocode to filter important sentences from answers}%
  \vspace{.5em}
  \Fn(){\FRecurs{$query,answer,pattern1,pattern2$}}{
  
   $removedSentences$ \\
   $stfCoreNLP$  \tcc*[f]{Stanford Core lib}\\
   $selectedSentences \gets answer.getBody()$\\
   $procBody \gets preProcess(selectedSentences)$\\
   $sentences \gets stfCoreNLP.getSentences(procBody)$\\

  \For{\textbf{each} $sentence$ \textbf{in} $sentences$}{
     $parseTree \gets sentence.constituencyParse()$\\
     $matcher1 \gets pattern1.matcher(parseTree)$\\
     $matcher2 \gets pattern2.matcher(parseTree)$\\
    
    \If{$(not(matcher1) \And not(matcher2))$}{ 
         \If{($not(specialSentence(sentence)$)}{ 
             $selectedSentences\gets selectedSentences.replace(sentence,\texttt{\char`\"})$\\
             
         }      	
     }
  }
  
     \Return $selectedSentences$\\
  
}

  \vspace{-1em}
\end{algorithm}

The algorithm receives four parameters: the pre-processed query, one recommended answer and the patterns. CROKAGE first performs standard natural language pre-processing on the body texts of the answer (line 5). Next, the algorithm annotates these processed texts using Stanford Part-Of-Speech Tagger (POS Tagger)~\cite{stanfordnlp} where each word of the sentence is assigned a POS tag (line 6). The algorithm then iterates over the sentences (lines 7 to 16). For each sentence, it builds the parse tree (line 8) and generates two pattern matchers to obtain the nodes that satisfy \textit{pattern1} and \textit{pattern2} (lines 9 and 10). If none of the two patterns are satisfied (line 11 to 15), the algorithm checks whether the sentence belongs to any special conditions (line 12). This special condition is composed of 4 heuristics. That is, the sentence does not contain: numbers, camel case words, important words (i.e., ``insert", ``replace", ``update")\footnote{the complete list of words is available at: https://bit.ly/2UxYgzc} and shared words with the query. If the sentence does not satisfy to any of these conditions, the algorithm removes the sentence from the selected sentences list (line 13). After iterating over all sentences, the algorithm returns only the valid sentences (\textit{selectedSentences} at line 17). If this list is not empty, it means that the answer contains valid sentences, which are returned as the explanation of associated code segment.

\begin{figure}[t]
\centerline{\includegraphics[width=0.5\textwidth]{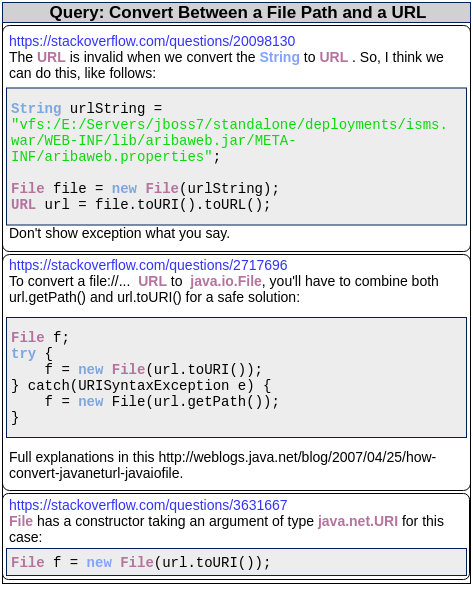}}
\vspace{-1.5em}
\caption{Comprehensive solution generated by CROKAGE for the query: \textit{``Convert Between a File Path and a URL"}}
\vspace{-1.5em}
\label{fig:summaryExample}
\end{figure}

Our intuition is that removing unnecessary sentences may help developers with a more concise explanation. CROKAGE is able to remove irrelevant sentences from the answers (e.g., \textit{``Try this:", ``You could do it like this:", ``It will work for sure", ``It seems the easiest to me"} or \textit{``Yes, like doing this"}). The output is a solution containing code and explanations. Figure~\ref{fig:summaryExample} shows Top-3 solutions comprising a comprehensive solution for the task (i.e., query): \textit{``Convert Between a File Path and a URL"}.

\section{Experiments and results} \label{experiments}

We evaluate our approach in different dimensions. First we evaluate the search for relevant answers (Section \ref{rankingMechanisms}) using 97 queries of our manually evaluated ground truth, of which 50\% was used for training and 50\% was used for testing. For the evaluation, we use four classical evaluation metrics and contrast the results with the six considered baselines, including the state-of-art BIKER~\cite{huang2018api}. Second, we perform a user study (Section \ref{userStudy}) with 29 developers using 24 queries to evaluate CROKAGE and BIKER in terms of relevance of the suggested code examples, benefit of the code explanations and the overall solution quality (code + explanation). In particular, we answer to three research questions using our experiments as follows:

\textbf{RQ1:} Can CROKAGE outperform existing baseline methods in retrieving relevant solutions for given programming tasks written as natural language queries?

\textbf{RQ2:} To what extent do the factors individually influence the ranking of candidate answers?

\textbf{RQ3:} Can CROKAGE provide more comprehensive solutions containing code and explanations for given queries (task descriptions) compared to those of the state-of-art, BIKER?

\begin{table*}[!ht]

\caption{CROKAGE parameters and their descriptions, ranges, variations and the highest values for Hit@10 and MRR@10.}
\vspace{-.5em}
\centering
\begin{tabular}{p{2.7cm}p{9.7cm}llc}
\hline
\textbf{Parameter}  & \textbf{Description}                                                                                                                                              & \multicolumn{1}{c}{\textbf{Range}} & \multicolumn{1}{c}{\textbf{Variation}} & \textbf{Best Value} \\ \hline
\textit{BM25Limit1}          & Top scored answers in BM25 to be used in semantic similarity relevance mechanism                       & {[}1k,10k{]}                       & 100                                    & 5k                  \\
\textit{BM25Limit2}          & Top scored answers in BM25 to be merged with the output of the semantic similarity relevance mechanism & {[}10,200{]}                       & 10                                     & 100                 \\
\textit{topAsymRelevanceNum} & Top scored answers by the semantic relevance mechanism   & {[}10,200{]}                       & 10                                     & 100                 \\
\textit{numberOfAPIClasses}  & Number of classes extracted from the three API Recommendation Systems combined                                                                                            & {[}5,30{]}                         & 5                                      & 20                  \\
\textit{semWeight}          & Weight associated with the semantic relevance score (semScore)                                                                                                 & {[}0,1{]}                          & 0.25                                   & 1.00                \\
\textit{apiWeight}           & Weight associated with the api score (apiScore)                                                                                                                   & {[}0,1{]}                          & 0.25                                   & 0.25                \\
\textit{tfidfWeight}        & Weight associated with TF-IDF score (tfidfScore)                                                                                                                      & {[}0,1{]}                          & 0.25                                   & 0.50                \\
\textit{methodWeight}        & Weight associated with method score (methodScore)                                                                                                                      & {[}0,1{]}                          & 0.25                                   & 0.75                \\ \hline
\end{tabular}
\label{parameters}
\vspace{-1em}
\end{table*}

\subsection{Ground Truth Generation} \label{groundTruth}

To build our ground truth, we first select 100 programming tasks (i.e., queries) from three Java tutorial sites: Java2s~\cite{java2s}, BeginnersBook~\cite{BeginnersBook} and KodeJava~\cite{KodeJava}. We select these queries in such a way so that they cover different API tasks and use these queries as input to three search engines: Google~\cite{Google}, Bing~\cite{Bing} and Stack Overflow search~\cite{SOSearch}. We pre-process each query by removing stop words, punctuation symbols, numbers and small words (length smaller then 2). For Google and Bing, we augment the query with the word \textit{``java"} (if the query does not contain it) and collect only such results that are from Stack Overflow. For Stack Overflow search, we filter results using the tag \textit{``java"}. We collect the first 10 results from Google and the first 20 from Bing. We observe lower efficiency of Stack Overflow search mechanism regarding the relevance of results when compared to Google and Bing. We thus establish a more rigorous criteria to fetch results from Stack Overflow search by setting a threshold of a minimum of 100 visualizations.

For each of the 100 queries, we merge results from the three search engines and remove duplicates. Results from these engines point to Stack Overflow threads. Each Stack Overflow thread is composed of a question and its answers. Since we are interested about the relevant answers to our query, we iterate over the questions, discard the question with no answers and select only answers with at least 1 upvote and containing source code. This automatic process results into 6,224 answers for 100 queries. Then, two professional developers manually evaluate the 6,224 answers by rating each answer in Likert scale from 1 to 5 according to the following criteria:

\textbf{1 =} Unrelated: the answer is not related to the query.

\textbf{2 =} Weakly related: the answer does not address the query problem objectively.

\textbf{3 =} Related: the answer needs considerable amount of changes in the source code to address the query problem, or is too long, or is too complex. 

\textbf{4 =} Understandable: the answer addresses the query problem after feasible amount of changes in the source code. 

\textbf{5 =} Straightforward: the answer addresses the query problem after few or no changes in the source code.

Two professional developers first evaluate the answers independently. After the evaluations, the average rating is calculated. If two ratings differ more than 1 Likert and at least one of them is higher than 3, this answer is marked to be re-evaluated by both in an agreement phase. The two professional developers then discuss these conflits. If after discussing, two ratings still differ in more than 1 Likert, this answer is discarded. We then re-calculate the average rating for the marked answers. We measure kappa before and after the agreement phase and we obtain the following values respectively: 0.3149 and 0.5063 (p-value $<$ 0.05). That is, our agreement improves from fair to moderate\cite{landis1977measurement}.

We consider an answer as relevant if its average Likert is equal or higher than 4. Since our approach relies on solutions containing API classes, we discard queries whose relevant answers do not contain API classes in their answers. Following this criteria, we discard three queries out of the 100, resulting in 97 queries along with their 1588 relevant answers.

\subsection{Performance Metrics} \label{performanceMetrics}

We choose four performance metrics commonly adopted by related literature~\cite{xu2017answerbot,rahman2018effective,rahman2016rack,rahman2017strict,huang2018api}. The four metrics are described as follows:

\textbf{Top-K Accuracy (Hit@K): } the percentage of search queries of which at least one recommended answer is relevant within the top-K results.

\textbf{Mean Reciprocal Rank (MRR@K): } the multiplicative inverse of the rank of the first relevant answer recommended within the top-K results.

\textbf{Mean Average Precision (MAP@K): } the average of all \textit{Precision@K} for a set of queries. \textit{Precision@K} is the precision of the all relevant answers within the first top-K recommnedations for every query.

\textbf{Mean Recall (MR@K): } the average of all \textit{Recall@K} for a set of queries. \textit{Recall@K} is the percentage of relevant answers recommended within the top-K results.

\subsection{Experimental Results for the Retrieval of Relevant Answers} \label{rankingMechanisms}

\textbf{RQ1:} \textit{Can CROKAGE outperform existing baseline methods in retrieving relevant solutions for given programming tasks written as natural language queries?}

We load the queries from our ground truth (i.e., 97 queries after discarding three), as well as their relevant answers namely \textit{goldSet} (i.e., average Likert equals or higher than 4). We split the queries into two sets namely training and test containing 49 and 48 queries respectively, along with their \textit{goldSets}. We use the training set and its \textit{goldSet} to calibrate the weights for each parameter of CROKAGE for which the Hit@K and MRR@K are the highest respectively. Table \ref{parameters} shows how we varied the parameters and the best values found for them. After discovering the best values for the parameters, we calibrate the weights and use the test set and its \textit{goldSet} to test CROKAGE and all the other baselines. To test CROKAGE, we run CROKAGE to search for relevant answers for each query of the test set (Section \ref{searchingforrelatedcontentinthecrowd}). We extend CROKAGE to generate the baselines as follows:

\textbf{BIKER:} BIKER extracts snippets from Stack Overflow answers to compose solutions. We extend the tool to show the answers IDs of which the snippets are extracted without altering its behaviour. 

\textbf{BM25:} we set CROKAGE to return only the top scored answers contained in the \textit{smallSet} (i.e., top \textit{lexicalSim},  Section \ref{bm25filter}, Fig.~\ref{fig:approach}-(C-1)).

\textbf{Semantic Relevance, API Class Relevance, TF-IDF Relevance and API Method Relevance:} we build four baselines representing CROKAGE relevance factors (Section~\ref{relevanceMechanism}, Fig.~\ref{fig:approach}-(C-4). For this, we preserve the weight associated to the baseline and set the other three weights (Formula~\ref{composer}) to zero (e.g., to build \textit{Semantic Relevance} baseline we set all factors' weights to zero, except \textit{semWeight}). 

After building all baselines, we run CROKAGE to search for relevant answers (Section \ref{searchingforrelatedcontentinthecrowd}) for each baseline against our test set. To evaluate each baseline, we compare their recommended answers against the \textit{goldSet} and collect the metrics Hit@K, MRR@K, MAP@K, and MR@K, for K=10 (i.e., we consider the top 10 recommendations). Table \ref{baselines} shows the metrics for all the baselines, including the state-of-art BIKER~\cite{huang2018api}. All the experiments were conducted over a server equipped with Intel\textregistered{} Xeon\textregistered{} at 3.1 GHz on 32 GB RAM, four cores, and 64-bit Linux Mint Cinnamon operating system. The total time to run the approach is around 188 seconds, of which 121 seconds are spend to load all models and indices and 67 seconds are spend to process the test set. That is, after loading the models, our approach takes less than 1.5 seconds to process each query.    

The non-parametric Wilcoxon signed-rank test on paired data showed significant difference between CROKAGE and BIKER for all considered metrics (i.e., p-values $<$ 0.05), with large effect size calculated with $r=Z/\sqrt{n}$ \cite{fritz2012effect} ranging from 0.61 to 0.82. Compared to BM25~\cite{robertson1994some}, this variation is lower: 17\%, 14\%, 8\% and 17\% for Hit@K, MAP@K, MR@K and MRR@K respectively (in absolute values), with a small-medium effect size (i.e., ranging from 0.25 to 0.35). 

\noindent
\fbox{\begin{minipage}{24em}
In terms of Top-K Accuracy, Mean Average Precision, Mean Recall, and Mean Reciprocal Rank for K=10, CROKAGE shows the highest values compared to all baselines, whereas BIKER~\cite{huang2018api} shows the lowest. For these metrics, CROKAGE significantly outperforms BIKER by 64\%, 30\%, 18\%, and 36\% respectively (in absolute values) to retrieve relevant answers for given programming tasks written in natural language (i.e., query). 

\end{minipage}}
\vspace{.5em}

\begin{table}[]
\caption{Performance of CROKAGE and other baseline methods in terms of Hit@K, MRR@K, MAP@K, and MR@K, for K=10} 
\centering
\begin{tabular}{lcccc}
\hline
Approach                   & Hit  & MRR  & MAP  & MR\\ \hline
BIKER                      & 0.15 & 0.10 & 0.10 & 0.01\\
API Class Relevance             & 0.38 & 0.11 & 0.11 & 0.06\\
API Method Relevance             & 0.46 & 0.17 & 0.15 & 0.06\\
Semantic Relevance       & 0.50 & 0.25 & 0.22 & 0.07\\
TF-IDF                     & 0.58 & 0.25 & 0.23 & 0.11\\
BM25                       & 0.62 & 0.29 & 0.26 & 0.11 \\
CROKAGE                    & 0.79 & 0.46 & 0.40 & 0.19 \\ \hline
\end{tabular}
\label{baselines}
\vspace{-1em}
\end{table}

\textbf{RQ2: }\textit{To what extent do the factors individually influence the ranking of candidate answers?}

CROKAGE obtains around 200 Stack Overflow candidate answers (i.e., duplicates are removed) to rank in the last stage of the search for relevant answers (Section~\ref{relevanceMechanism}). We investigate the individual influence of each of the four factors on candidate answers ranking in terms of Top-K Accuracy, Mean Reciprocal Rank, Mean Average Precision and Mean Recall, for K=10 (i.e., considering top 10 recommendations). For this, we extend CROKAGE by setting the weight associated to each factor to zero (Formula~\ref{composer}), while keeping the others as follows: 

\textbf{CROKsemWeight0}: we set \textit{semWeight} to zero, representing the semantic relevance influence.

\textbf{CROKapiWeight0}: we set \textit{apiWeight} to zero, representing the API Class relevance influence.

\textbf{CROKtfidfWeight0}: we set \textit{tfidfWeight} to zero, representing TF-IDF relevance influence.

\textbf{CROKmethodWeight0}: we set \textit{methodWeight} to zero, representing API Method relevance influence.

We run each version against our ground truth queries to search for relevant answers (Section \ref{searchingforrelatedcontentinthecrowd}) and collect the metrics as shown in Table~\ref{tab:crokageversions}. 

\vspace{1em}
\noindent
\fbox{\begin{minipage}{24em}
Our findings suggest that, in the presence of the other factors associated with their calibrated weights, TF-IDF shows the highest influence on the ranking of candidate answers. Compared to CROKAGE, the version with TF-IDF weight (tfidfWeight) set to zero shows values for metrics 8\%, 15\%, 13\%, and 8\% lower for Top-K Accuracy, Mean Reciprocal Rank, Mean Average Precision, and Mean Recall respectively. This difference is also significant for API Method (i.e., 14\% in Top-K Accuracy). Semantic and API Class relevance instead, show small influence on the ranking of candidate answers (i.e., 8\% and 4\% in Mean Reciprocal Rank respectively).
\end{minipage}}
\vspace{.5em}

In order to collect the API classes for a given query (i.e., task description), we use three state-of-art API recommendation systems -- BIKER~\cite{huang2018api}, NLP2API~\cite{rahman2018effective} and RACK~\cite{rahman2016rack}. We investigate their combination to provide API classes in such a way to obtain the highest Top-K Accuracy, Mean Reciprocal Rank, Mean Average Precision and Mean Recall, respectively. We tested 308 queries using the dataset and ground truth of NLP2API in terms of the metrics. In general, the combination of the three tools performs better than two tools combined or each tool taken isolated for different values of K (e.g., 1, 5, 10). We employ the combination of RACK + BIKER + NLP2API to provide API classes to a given query considering the order of the recommendations. We tested different numbers of API classes (Table \ref{parameters}) and found that collecting 20 classes from the three approaches combined gives the best performance. We show the metrics for each tool and for their combination with the highest performance for K=10 (i.e., top 10 recommendations) in Table~\ref{tab:nlp2api}.

\begin{table}[]
\centering
\caption{Performance of four extended versions of CROKAGE in terms of Hit@K, MRR@K, MAP@K, and MR@K, for K=10}
\vspace{-1em}
\begin{tabular}{lcccc}
\hline
Version     & Hit  & MRR  & MAP  & MR   \\ \hline
CROKtfidfWeight0 & 0.71 & 0.31 & 0.27 & 0.11 \\
CROKmethodWeight0  & 0.65 & 0.37 & 0.31 & 0.14 \\
CROKsemWeight0    & 0.79 & 0.38 & 0.36 & 0.19 \\
CROKapiWeight0   & 0.83 & 0.42 & 0.37 & 0.18 \\
 \hline
\end{tabular}
\label{tab:crokageversions}
\vspace{-2em}
\end{table}

\begin{table}[!ht]
\vspace{-1.5em}
\centering
\caption{Performance of three API Recommendation Systems and two different combinations in nlp2api dataset in terms of Hit@K, MRR@K, MAP@K, and MR@K, for K=10}
\vspace{-1em}
\begin{tabular}{lcccc}
\hline
Approach     & Hit  & MRR  & MAP  & MR   \\ \hline
RACK (RA)    & 0.74 & 0.47 & 0.42 & 0.40 \\
BIKER (BI)   & 0.52 & 0.37 & 0.35 & 0.23 \\
NLP2API (NL) & 0.73 & 0.49 & 0.44 & 0.38 \\
RA+NL        & 0.80 & 0.50 & 0.44 & 0.46 \\
RA+BI+NL     & 0.83 & 0.51 & 0.44 & 0.47 \\ \hline
\end{tabular}
\label{tab:nlp2api}
\vspace{-1.5em}
\end{table}

\subsection{Comparison with State-of-the-art using Developer Study} \label{userStudy}

\textbf{RQ3:} \textit{Can CROKAGE provide more comprehensive solutions containing code and explanations for given queries (task descriptions) compared to those of the state-of-art, BIKER?}

To answer this question, we first choose 50 most popular questions from the same three tutorial sites used to generate our ground truth (discarding questions already used to build the ground truth). We augment the question with the filter ``\textit{site:stackoverflow.com}" and use Google to measure the popularity of each one. For this, we check the number of Google's results returned for them. We randomly select 30 among these questions and apply standard natural language pre-processing. We assume that developers would use CROKAGE like a search engine. Thus, we restrict the queries to not contain too many words (i.e., maximum of seven and a minimum of two tokens). This process results in 24 queries. We run BIKER~\cite{huang2018api} and CROKAGE to generate the solutions for these queries. We focus in providing solutions with high precision. Thus, we set both tools to use only the top 1 recommended answer to build solutions. We then ask developers to evaluate the tools in terms of three aspects: 

\begin{enumerate}
\item The relevance of the suggested code examples
\item The benefit of the code explanations
\item The overall solution quality (code + explanation) 
\end{enumerate}

\begin{figure*}[!ht]
\centering
\subfloat[Relevance of the Suggested Code Examples]{
    \scalebox{0.5}{
    \begin{tikzpicture}
     \begin{axis}
        [
        boxplot/draw direction=y,
        xtick={1,2,3,4,5},
        xticklabels={A, B},
        yticklabel style = {font=\Large},
        xticklabel style = {font=\Large}
        ]
        \addplot[
        boxplot prepared={
         median=3,
         upper quartile=4,
         lower quartile=1.05,
         upper whisker=5,
         lower whisker=1
        },
        ] coordinates {};
        \addplot[
        boxplot prepared={
         median=4,
         upper quartile=4.95,
         lower quartile=3,
         upper whisker=5,
         lower whisker=1
        },
        ] coordinates {};
     \end{axis}
    \end{tikzpicture}
    \label{fig_code}
    }
}
\hfil
\subfloat[Benefit of the Code
Explanations]{
    \scalebox{0.5}{
    \begin{tikzpicture}
     \begin{axis}
        [
        boxplot/draw direction=y,
        xtick={1,2,3,4,5},
        xticklabels={A, B},
        yticklabel style = {font=\Large},
        xticklabel style = {font=\Large}
        ]
        \addplot[
        boxplot prepared={
         median=3,
         upper quartile=4,
         lower quartile=1.05,
         upper whisker=5,
         lower whisker=1
        },
        ] coordinates {};
        \addplot[
        boxplot prepared={
         median=3,
         upper quartile=4,
         lower quartile=2,
         upper whisker=5,
         lower whisker=1
        },
        ] coordinates {};
     \end{axis}
    \end{tikzpicture}
    \label{fig_explan}
    }
}
\hfil
\subfloat[Overall Solution Quality]{
    \scalebox{0.5}{
    \begin{tikzpicture}
     \begin{axis}
        [
        boxplot/draw direction=y,
        xtick={1,2},
        xticklabels={A, B},
        yticklabel style = {font=\Large},
        xticklabel style = {font=\Large}
        ]
        \addplot[draw=black,mark=*,mark options = {mark color=black},
        boxplot prepared={
         median=3,
         upper quartile=4,
         lower quartile=1.05,
         upper whisker=5,
         lower whisker=1
        },
        ] coordinates {};
        \addplot[draw=black,mark=*,mark options = {mark color=black},
        boxplot prepared={
         median=4,
         upper quartile=4.05,
         lower quartile=3,
         upper whisker=5,
         lower whisker=2
        },
        ] coordinates{(0,1)};
     \end{axis}
    \end{tikzpicture}
    \label{fig_overall}
    }
}
\caption{Box plots of Relevance of the Suggested Code Examples (a), Benefit of the Code
Explanations (b), and the Overall Solution Quality (c) performance (Likert scale) for tools A (BIKER) and B (CROKAGE). Lower and upper box boundaries 25\textsuperscript{th} and 75\textsuperscript{th} percentiles, respectively, line inside box median. Lower and upper error lines 10\textsuperscript{th} and 90\textsuperscript{th} percentiles, respectively. Filled circle data falls outside 10\textsuperscript{th} and 90\textsuperscript{th} percentiles.}
\vspace{-0.9em}
\label{fig:boxplot}
\end{figure*}
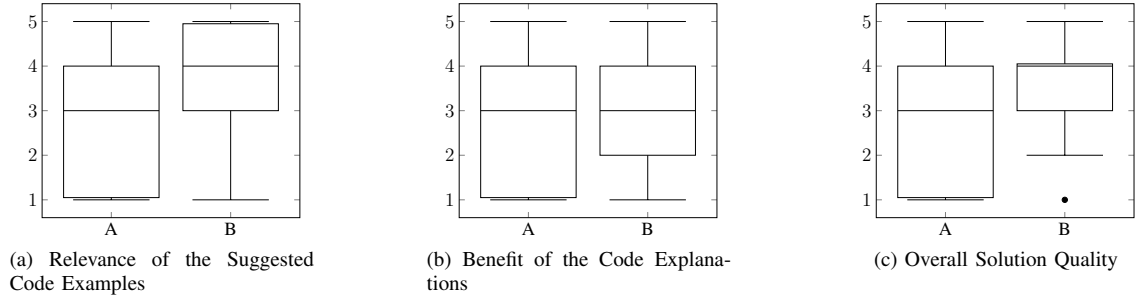

We built three questionnaires, each containing eight different questions and their solutions for both tools (identified as Tool A and Tool B). We also provided instructions for the evaluations. We asked participants to provide a value from 1 to 5 for each aspect in each tool considering the same criteria used to evaluate our ground truth (Section \ref{groundTruth}). We also asked the participants how many years of experience they have as Java programmers and as Professional Software developers. The averages are 7.92 and 8.78 years respectively. In total, 29 participants answered our study and each query was answered by at least nine participants. We assured that each questionnaire contains at least five participants with a minimum of six years of experience in Java programming.

Figure~\ref{fig:boxplot} represents the participants evaluations. In general, participants reported that Tool B (CROKAGE) was considered better than Tool A (BIKER~\cite{huang2018api}). For the three considered criteria, two of them were reported by users as much superior: relevance of the suggested code examples (Fig.~\ref{fig:boxplot}-(a)) and the overall solution quality (Fig.~\ref{fig:boxplot}-(c)). Both approaches showed the same median value (i.e,. 3) for the benefit of the code explanations (Fig.~\ref{fig:boxplot}-(b)). However CROKAGE showed much less Likerts 1 (i.e., 19 against 72) and much more Likerts 5 (i.e., 52 against 20). These results suggest that developers prefer explanations provided by other users instead of generic explanations extracted from official API documentations. 

\vspace{.5em}
\noindent
\fbox{\begin{minipage}{24em}
Our findings show that CROKAGE outperforms BIKER\cite{huang2018api} for the three considered criteria: relevance of the suggested code examples (Fig.~\ref{fig:boxplot}-(a)), benefit of the code explanations (Fig.~\ref{fig:boxplot}-(b)) and the overall solution quality (Fig.~\ref{fig:boxplot}-(c)).
\end{minipage}}
\vspace{.5em} 

To make sure that the performance of CROKAGE and BIKER are statistically different, we run Wilcoxon signed-rank~\cite{wilcoxon1945individual} for each adopted criteria. We find that the evaluation of CROKAGE is statistically better than the one of BIKER for the three criteria with a confidence level of 95\% (p-value $< 0.05$) and a medium effect size ranging from 0.40 to 0.42, calculated with $r=Z/\sqrt{n}$  \cite{fritz2012effect}.

\section{Threats to validity}\label{sec:threats}

\textbf{Threats to internal validity} are related to the baseline methods and the user study. One of the baselines is a state-of-art tool and we extend it to produce the solutions with supplementary information (i.e., answers IDs) without altering its behavior. We double checked the implementation of all baselines to assure they do not contain implementation errors. For the user study, the experience of each participant in Java programming and their effort in manual evaluation could affect the accuracy of the results. We mitigate this threat by organizing questionnaires where each questionnaire has at least five participants with a minimum of six years of experience in Java programming. We only selected participants who showed interest in participating in our research.

\textbf{Threats to external validity} relates to the quality of our ground truth and to the generalizability of our results. Concerning the ground truth, to mitigate the threat two professional developers independently evaluated each answer and then discussed the disagreements. Furthermore, we only selected good quality answers for the \textit{goldSet} (i.e., Likert equal or higher than 4). We tried to cover a wide range of different API tasks by selecting programming tasks from three popular tutorial sites. CROKAGE only considers Java-tagged questions and their answers due to the background knowledge of the evaluators. However, we consider this as a limitation instead of a threat, since our dataset contains 3.8 M of Java posts. CROKAGE could be easily adapted to support recommendations to other languages as long as it can extract information from Stack Overflow. 

\textbf{Threats to construct validity} relates to suitability of our evaluation metrics. We choose four performance metrics: Top-K Accuracy, Mean Reciprocal Rank, Mean Average Precision and Mean Recall, which are commonly adopted by related literature in software engineering\cite{xu2017answerbot,rahman2018effective,rahman2016rack,rahman2017strict,huang2018api}.

\section{Related Work}\label{relatedWord}

\subsection{Code Example Suggestion}

There have been several studies~\cite{nguyen2016t2api,bajracharya2010searching,6233407,campbell2017nlp2code,wang2016hunter,gvero2015interactive,raghothaman2016swim,gu2016deep,campos2014nuggets,mcmillan2011portfolio,rahman2017rack,gu2018deep,huang2018api} that return relevant code against natural language queries. McMillan~et~al.~\cite{mcmillan2011portfolio} propose a search engine that combines PageRank with Keyword matching to retrieve relevant functions. Our work differs from theirs on the granularity of the suggested code, since we do not restrict our search to functions. Campbell~and~Treude~\cite{campbell2017nlp2code} develop a tool that assists users providing suggestions to the queries. Rahman~et~al.~\cite{rahman2017rack} instead, propose a tool to reformulate the queries before applying the search by using associations between keywords and APIs. Both tools however rely on third-part search engines. While the first relies on Google search API to retrieve relevant code, the second uses GitHub code search API. This dependency constrains their tools to the limitations of the third-part APIs (e.g., the number of searches in a period of time).  

Some works~\cite{nguyen2016t2api,gu2016deep,raghothaman2016swim} infer API usage sequences for a given task. T2API~\cite{nguyen2016t2api} learns API usages via graph-based language model. They use a statistical machine translation to associate descriptions and corresponding code. DeepAPI~\cite{gu2016deep} composes the associations between the sequence of words in a query and APIs through deep learning. SWIM~\cite{raghothaman2016swim} uses statistical word alignment to relate query words with API elements. Our work instead, exploits more than just API sequences. While these tools could return the same API sequences for two queries with different purposes, our work distinguish code aspects like method and class names. This concern has been also addressed by DeepCS~\cite{gu2018deep}. Their tool jointly embeds natural language descriptions and code examples into a high-dimensional vector space in such a way that the description and their accompanying code examples have similar vector representations. They use such representations to calculate the similarity between the query and the code. Our tool is similar, but instead of using deep learning, we rely on information retrieval techniques. 

Several tools~\cite{campos2014nuggets,lv2015codehow,6233407,bajracharya2010searching,7884618,Dotzler:2017} rely on lexical similarities to retrieve relevant code. Campos~et~al.~\cite{campos2014nuggets} rank related code documents by applying a combination of Vector Space and Boolean models. The same idea is used by Lv~et~al.~\cite{lv2015codehow}. Their tool however, extends the Boolean model to integrate the benefits of both models. Like our tool, Lv~et~al.'s tool also enriches the search with API names related to the input query. Zagalsky~et~al.~\cite{6233407} propose a tool to retrieve source code based on keywords using TF-IDF to score code documents. Bajracharya~et~al.~\cite{bajracharya2010searching} mine relevant API elements through shared concepts between the query and suggested words from open source systems. These mentioned approaches however, miss relevant documents if the query and the documents do not share common words. Our tool addresses this weakness by harnessing embeddings to capture words' semantics. That is, our tool is able to find documents that share semantically similar words with the query, despite having lexical dissimilarity. Furthermore, our tool can distinguish the order of the words, another limitation of their approaches.

Our work, differently from the mentioned tools, not only retrieves code but also provides explanations. 
BIKER~\cite{huang2018api} is the most related work to ours and we compare our work with theirs in multiple ways, as shown in Section \ref{experiments}.

\subsection{Code Explanation Generation}
Several early studies~\cite{wong2013autocomment,rahman2015recommending,xu2017answerbot,deSouza:2014:RCK,campos2016searching,chatterjee2017extracting,hu2018deep,wong2015clocom} propose automatic approaches to extract explanations to code. For this, they explore lexical properties usually in combination with strategies like clone detection~\cite{wong2013autocomment,wong2015clocom}, topics (like LDA)~\cite{rahman2015recommending}, word embeddings~\cite{xu2017answerbot}, machine learning~\cite{chatterjee2017extracting} and deep learning~\cite{hu2018deep}. Wong~et~al.~\cite{wong2013autocomment} propose a series of heuristics to match the code with natural language. They select the best descriptions for a code and use natural language processing to filter relevant sentences to compose the descriptions. Our work harness two patterns they develop to select relevant sentences. Similarly, in another study, Wong~et~al.~\cite{wong2015clocom} synthesize comments from similar code snippets. They try to address the limitation of their previous work by using GitHub instead of Stack Overflow to extract the comments, since comments in Q\&A websites are not often written in full sentences. Rahman~et~al.~\cite{rahman2015recommending} use heuristics to extract comments from Stack Overflow. Their approach combines the heuristics to rank the top most relevant comments for a source code. Chatterjee~et~al.~\cite{chatterjee2017extracting} develop a technique to extract descriptions associated with code segments from articles. Differently from Q\&A websites, the code in articles is not delineated by markers. They also convert documents (e.g., pdf and images) to text and learn the associations between text and code using machine learning. Xu~et~al.~\cite{xu2017answerbot} employ word embeddings to handle the lexical gap between between natural language queries and Stack Overflow question titles. They use the answers from relevant questions to produce summaries. Despite they generate diverse summaries to the queries, their summaries do not contain source code. Hu~et~al.~\cite{hu2018deep} propose an approach to generate comments for java methods through neural networks. But instead of relying on words to learn associations between code and descriptions, they use Abstract Syntax Trees to represent  methods. This strategy showed efficiency to lean the associations even when methods and identifiers in the code are poorly named. 

We refer the reader to the survey by Wang~et~al.~\cite{wang2018comment} to more information about works in the context of comment generation for source code. Our work is closely related to these works in the sense that we also capture explanations for source code. We leverage natural language processing and explore lexical properties by considering the context surrounding the code. 

\section{Conclusion and Future Work} \label{conclusion}

In this work, we propose CROKAGE, a tool to help developers with the daily problem of seeking relevant code examples on the web for programming tasks. CROKAGE leverages API knowledge stored in Stack Overflow to generate solutions containing source code and explanations for tasks written in natural language. For this, we first employ lexical similarity combined with word embeddings to select candidate answers from Stack Overflow to a programming task (i.e., query). Then, we assign on each answer an API class score according to three state-of-art API recommendation systems. Next, we combine four weighted factors to rank candidate answers. And lastly, we use natural language processing on the top quality answers to compose the solutions. Our findings show that CROKAGE outperforms several other baselines to retrieve relevant answers for programming tasks, including a state-of-the-art one. We also perform a user study that demonstrate the effectiveness of CROKAGE to provide quality solutions regarding code examples and explanations. In the future, we will implement CROKAGE in form of an Eclipse plugin to enable developers to obtain our solutions right from the IDE. 

\section*{Acknowledgement}
We thank the authors of BIKER for sharing their tool. This research is supported in-part by a Canada First Research Excellence Fund (CFREF) grant coordinated by the Global Institute for Food Security (GIFS).

\bibliographystyle{IEEEtran}
\bibliography{biblio} 

\begin{thebibliography}{10}
\providecommand{\url}[1]{#1}
\csname url@samestyle\endcsname
\providecommand{\newblock}{\relax}
\providecommand{\bibinfo}[2]{#2}
\providecommand{\BIBentrySTDinterwordspacing}{\spaceskip=0pt\relax}
\providecommand{\BIBentryALTinterwordstretchfactor}{4}
\providecommand{\BIBentryALTinterwordspacing}{\spaceskip=\fontdimen2\font plus
\BIBentryALTinterwordstretchfactor\fontdimen3\font minus
  \fontdimen4\font\relax}
\providecommand{\BIBforeignlanguage}[2]{{%
\expandafter\ifx\csname l@#1\endcsname\relax
\typeout{** WARNING: IEEEtran.bst: No hyphenation pattern has been}%
\typeout{** loaded for the language `#1'. Using the pattern for}%
\typeout{** the default language instead.}%
\else
\language=\csname l@#1\endcsname
\fi
#2}}
\providecommand{\BIBdecl}{\relax}
\BIBdecl

\bibitem{rahman2018effective}
M.~M. Rahman and C.~K. Roy, ``Effective reformulation of query for code search
  using crowdsourced knowledge and extra-large data analytics,'' in \emph{Proc.
  ICSME}, 2018, pp. 473--484.

\bibitem{xu2017answerbot}
B.~Xu, Z.~Xing, X.~Xia, and D.~Lo, ``Answerbot: Automated generation of answer
  summary to developers\'{z} technical questions,'' in \emph{Proc. ASE}, 2017,
  pp. 706--716.

\bibitem{mikolov2013distributed}
T.~Mikolov, I.~Sutskever, K.~Chen, G.~Corrado, and J.~Dean, ``Distributed
  representations of words and phrases and their compositionality,'' in
  \emph{Proc. NIPS}, 2013, pp. 3111--3119.

\bibitem{huang2018api}
Q.~Huang, X.~Xia, Z.~Xing, D.~Lo, and X.~Wang, ``{API} method recommendation
  without worrying about the task-{API} knowledge gap,'' in \emph{Proc. ASE},
  2018, pp. 293--304.

\bibitem{salton1988term}
G.~Salton and C.~Buckley, ``Term-weighting approaches in automatic text
  retrieval,'' \emph{IP\&M}, vol.~24, no.~5, pp. 513--523, 1988.

\bibitem{robertson1994some}
S.~E. Robertson and S.~Walker, ``Some simple effective approximations to the
  2-poisson model for probabilistic weighted retrieval,'' in \emph{Proc. ACM
  SIGIR}, 1994, pp. 232--241.

\bibitem{Jsoup}
Jsoup, ``Java html parser,'' \url{http://jsoup.org}.

\bibitem{delfim2016redocumenting}
F.~M. Delfim, K.~V.~R. {Paixao}, D.~Cassou, and M.~A. de~Almeida~Maia,
  ``Redocumenting apis with crowd knowledge: a coverage analysis based on
  question types,'' \emph{JBCS}, vol.~22, no.~1, p.~9, 2016.

\bibitem{Lucene}
Apache, ``Lucene,'' \url{http://lucene.apache.org/}.

\bibitem{bojanowski2016enriching}
P.~Bojanowski, E.~Grave, A.~Joulin, and T.~Mikolov, ``Enriching word vectors
  with subword information,'' \emph{TACL}, vol.~5, pp. 135--146, 2017.

\bibitem{hill2012use}
E.~{Hill}, S.~{Rao}, and A.~{Kak}, ``On the use of stemming for concern
  location and bug localization in java,'' in \emph{Proc. SCAM}, 2012, pp.
  184--193.

\bibitem{ye2016word}
X.~Ye, H.~Shen, X.~Ma, R.~Bunescu, and C.~Liu, ``From word embeddings to
  document similarities for improved information retrieval in software
  engineering,'' in \emph{Proc. ICSE}, 2016, pp. 404--415.

\bibitem{Ahasanuzzaman:2016}
M.~Ahasanuzzaman, M.~Asaduzzaman, C.~K. Roy, and K.~A. Schneider, ``Mining
  duplicate questions in stack overflow,'' in \emph{Proc. MSR}, 2016, pp.
  402--412.

\bibitem{silva2018duplicate}
R.~F.~G. {Silva}, K.~V.~R. {Paixao}, and M.~A. Maia, ``Duplicate question
  detection in stack overflow: A reproducibility study,'' in \emph{Proc.
  SANER}, 2018, pp. 572--581.

\bibitem{rahman2016rack}
M.~M. Rahman, C.~K. Roy, and D.~{Lo}, ``Rack: Automatic api recommendation
  using crowdsourced knowledge,'' in \emph{Proc. SANER}, 2016, pp. 349--359.

\bibitem{wong2013autocomment}
E.~Wong, J.~Yang, and L.~Tan, ``Autocomment: Mining question and answer sites
  for automatic comment generation,'' in \emph{Proc. ASE}, 2013, pp. 562--567.

\bibitem{stanfordnlp}
C.~D. Manning, M.~Surdeanu, J.~Bauer, J.~Finkel, S.~J. Bethard, and
  D.~McClosky, ``The {Stanford} {CoreNLP} natural language processing
  toolkit,'' in \emph{Proc. ACL}, 2014, pp. 55--60.

\bibitem{java2s}
Java2s, ``Java2s,'' \url{http://java2s.com}.

\bibitem{BeginnersBook}
BeginnersBook, ``{BeginnersBook},'' \url{http://beginnersbook.com}.

\bibitem{KodeJava}
W.~Saryada, ``Kodejava,'' \url{http://kodejava.org}.

\bibitem{Google}
{Google Inc.}, ``Google search engine,'' \url{http://google.com}.

\bibitem{Bing}
{Microsoft Inc.}, ``Bing search engine,'' \url{http://bing.com}.

\bibitem{SOSearch}
{Stack Exchange Inc.}, ``{Stack Overflow} search engine,''
  \url{http://stackoverflow.com}.

\bibitem{landis1977measurement}
J.~R. Landis and G.~G. Koch, ``The measurement of observer agreement for
  categorical data,'' \emph{Biometrics}, vol.~33, no.~1, pp. 159--174, 1977.

\bibitem{rahman2017strict}
M.~M. Rahman and C.~K. Roy, ``{STRICT}: Information retrieval based search term
  identification for concept location,'' in \emph{Proc. SANER}, 2017, pp.
  79--90.

\bibitem{fritz2012effect}
C.~Fritz, E.~Peter, and J.~Richler, ``Effect size estimates: current use,
  calculations, and interpretation.'' \emph{JEPG}, vol. 141, no.~1, p. 2–18,
  2012.

\bibitem{wilcoxon1945individual}
F.~Wilcoxon, ``Individual comparisons by ranking methods,'' \emph{Biometrics
  bulletin}, vol.~1, no.~6, pp. 80--83, 1945.

\bibitem{nguyen2016t2api}
T.~Nguyen, P.~C. Rigby, A.~T. Nguyen, M.~Karanfil, and T.~N. Nguyen, ``{T2API}:
  synthesizing {API} code usage templates from english texts with statistical
  translation,'' in \emph{Proc. FSE}, 2016, pp. 1013--1017.

\bibitem{bajracharya2010searching}
S.~Bajracharya, J.~Ossher, and C.~Lopes, ``Searching {API} usage examples in
  code repositories with sourcerer {API} search,'' in \emph{Workshop on
  Search-driven Development}, 2010, pp. 5--8.

\bibitem{6233407}
A.~Zagalsky, O.~Barzilay, and A.~Yehudai, ``{Example Overflow}: Using social
  media for code recommendation,'' in \emph{Proc. RSSE}, 2012, pp. 38--42.

\bibitem{campbell2017nlp2code}
B.~A. {Campbell} and C.~{Treude}, ``Nlp2code: Code snippet content assist via
  natural language tasks,'' in \emph{Proc. ICSME}, 2017, pp. 628--632.

\bibitem{wang2016hunter}
Y.~Wang, Y.~Feng, R.~Martins, A.~Kaushik, I.~Dillig, and S.~P. Reiss, ``Hunter:
  next-generation code reuse for java,'' in \emph{Proc. FSE}, 2016, pp.
  1028--1032.

\bibitem{gvero2015interactive}
T.~Gvero and V.~Kuncak, ``Interactive synthesis using free-form queries,'' in
  \emph{Proc. ICSE}, 2015, pp. 689--692.

\bibitem{raghothaman2016swim}
M.~Raghothaman, Y.~Wei, and Y.~Hamadi, ``{SWIM}: Synthesizing what {I}
  mean-code search and idiomatic snippet synthesis,'' in \emph{Proc. ICSE},
  2016, pp. 357--367.

\bibitem{gu2016deep}
X.~Gu, H.~Zhang, D.~Zhang, and S.~Kim, ``Deep {API} learning,'' in \emph{Proc.
  FSE}, 2016, pp. 631--642.

\bibitem{campos2014nuggets}
E.~C. Campos, L.~B. L.~D. Souza, and M.~A. Maia, ``Nuggets miner: Assisting
  developers by harnessing the {Stack Overflow} crowd knowledge and the github
  traceability,'' in \emph{Proc. CBSoft-Tool Session}, 2014.

\bibitem{mcmillan2011portfolio}
C.~McMillan, M.~Grechanik, D.~Poshyvanyk, Q.~Xie, and C.~Fu, ``Portfolio:
  finding relevant functions and their usage,'' in \emph{Proc. ICSE}, 2011, pp.
  111--120.

\bibitem{rahman2017rack}
M.~M. Rahman, C.~K. Roy, and D.~Lo, ``Rack: Code search in the {IDE} using
  crowdsourced knowledge,'' in \emph{Proc. ICSE}, 2017, pp. 51--54.

\bibitem{gu2018deep}
X.~Gu, H.~Zhang, and S.~Kim, ``Deep code search,'' in \emph{Proc. ICSE}, 2018,
  pp. 933--944.

\bibitem{lv2015codehow}
F.~Lv, H.~Zhang, J.-g. Lou, S.~Wang, D.~Zhang, and J.~Zhao, ``Codehow:
  Effective code search based on {API} understanding and extended boolean model
  (e),'' in \emph{Proc. ASE}, 2015, pp. 260--270.

\bibitem{7884618}
G.~{Santos}, K.~V.~R. {Paixao}, N.~{Anquetil}, A.~{Etien}, M.~A. Maia, and
  S.~{Ducasse}, ``Recommending source code locations for system specific
  transformations,'' in \emph{Proc. SANER}, 2017, pp. 160--170.

\bibitem{Dotzler:2017}
G.~Dotzler, M.~Kamp, P.~Kreutzer, and M.~Philippsen, ``More accurate
  recommendations for method-level changes,'' in \emph{Proc. ESEC/FSE}, 2017,
  pp. 798--808.

\bibitem{rahman2015recommending}
M.~M. Rahman, C.~K. Roy, and I.~Keivanloo, ``Recommending insightful comments
  for source code using crowdsourced knowledge,'' in \emph{Proc. SCAM}, 2015,
  pp. 81--90.

\bibitem{deSouza:2014:RCK}
L.~B.~L. de~Souza, E.~C. Campos, and M.~A. Maia, ``Ranking crowd knowledge to
  assist software development,'' in \emph{Proc. ICPC}, 2014, pp. 72--82.

\bibitem{campos2016searching}
E.~C. Campos, L.~B. de~Souza, and M.~A. Maia, ``Searching crowd knowledge to
  recommend solutions for api usage tasks,'' \emph{JSEP}, vol.~28, no.~10, pp.
  863--892, 2016.

\bibitem{chatterjee2017extracting}
P.~Chatterjee, B.~Gause, H.~Hedinger, and L.~Pollock, ``Extracting code
  segments and their descriptions from research articles,'' in \emph{Proc.
  MSR}, 2017, pp. 91--101.

\bibitem{hu2018deep}
X.~Hu, G.~Li, X.~Xia, D.~Lo, and Z.~Jin, ``Deep code comment generation,'' in
  \emph{Proc. ICPC}, 2018, pp. 200--210.

\bibitem{wong2015clocom}
E.~Wong, T.~Liu, and L.~Tan, ``Clocom: Mining existing source code for
  automatic comment generation,'' in \emph{Proc. SANER}, 2015, pp. 380--389.

\bibitem{wang2018comment}
X.~Wang, Y.~Peng, and B.~Zhang, ``Comment generation for source code: State of
  the art, challenges and opportunities,'' \emph{arXiv:1802.02971}, 2018.

\end{thebibliography}

\end{document}